\documentclass[]{llncs}

\pdfoutput=1

\usepackage{latexsym}
\usepackage{amsfonts}
\usepackage{amsmath}
\usepackage{listings}

\usepackage{url}
\usepackage{graphicx}
\usepackage{subfigure}

\newcommand{\remove}[1]{}
\newcommand{\HVE}{{\mathsf{HVE}}}

\newcommand{\comment} [1]{}                             						

\newcommand{\x}{{\vec{x}}}

\newcommand{\G}{\mathbb{G}}

\newcommand{\F}{\mathbb{F}}
\newcommand{\e}{\mathbf{e}}

\title {Answering queries using pairings}
\author{Alberto Trombetta\inst{1} \and Giuseppe Persiano\inst{2} \and Stefano Braghin\inst{3}}
\institute{University of Insubria, DiSTA, Varese, Italy\\\email{alberto.trombetta@uninsubria.it}
\and
University of Salerno, DIA, Salerno, Italy\\\email{giuper@dia.unisa.it}
\and
IBM Research -- Ireland, Smarter Cities Technology Centre, Dublin, Ireland\\\email{stefanob@ie.ibm.com}}
 
\begin{document}

\maketitle

\begin{abstract}
Outsourcing data in the cloud has become nowadays very common.
Since -- generally speaking -- cloud data storage and management 
providers cannot be fully trusted, mechanisms providing the confidentiality of
the stored data are necessary.
A possible solution is to encrypt all the data, but -- of course --
this poses serious problems about the effective usefulness of the stored data.
In this work, we propose to apply a well-known attribute-based cryptographic
scheme to cope with the problem of querying encrypted data.
We have implemented the proposed scheme with a real-world, off-the-shelf RDBMS
and we provide several experimental results showing the feasibility of our approach.
 
\end{abstract}

\section{Introduction} 
\label{sec:intro}
The widespread and ever-growing deployment of cloud computing allows users to access a wide array of services, such as online data storage. The
advantages in adopting such solutions is mitigated by the fact that outsourced data is not under the direct control of the legitimate owner, who has
uploaded it to the storage service and one has to fully trust the service provider.
As a typical solution, the data owner encrypts with a classical symmetric scheme its data before uploading it to the storage provider.
This is not optimal in the case the storage manager -- or some other party other than the data owner -- has to perform some kind of computation
over the encrypted data. For example, it is impossible to search for a given content over encrypted data, or,
more specifically, in the case of encrypted structured data (such as relational data), it is not possible to perform expressive, SQL-like queries.
Several solutions have been proposed that solve the problem of searching over encrypted data. However, most of such solutions are not well-tailored for performing searches over structured data, such as relational tables. We refer to Section \ref{sec:rel} for a discussion.

\paragraph{Our motivating scenario.} Consider a data owner $U$ that collects data coming from an information source (say, a 
sensor network) and stores them in a cloud-based database, managed by a (possibly untrusted) third party. We assume that the outsourced database
is formed by a single table $T$. In the following we describe the security requirements related to the collection and querying of such data.
The information collected in the table $T$ is sensitive and should not be 
unconditionally disclosed. 
Henceforth, in this scenario the solutions typically 
adopted for
ensuring the confidentiality (and then enforcing proper access policies to the data) do not apply. In fact, the DBMS is typically demanded to
enforce the access control to the data stored in it, but -- since in our case the DBMS could be untrusted -- this solution is not viable.
The first few rows of an instance of the table $T$ are as follows:

{\tiny
\begin{center}
\begin{tabular}{|c|c|c|c|c|c|c|}
\hline
ServiceId & TypeId & Availability & Certificate & Position & Description & Timestamp\\
\hline
\hline 
10 & 3 & yes & $cert_1$ & District2 & infrared camera & 2010-09-07 12:45:33\\
23 & 3 & yes & $cert_2$ & District3 & temperature & 2012-12-03 11:52:34\\
41 & 1 & yes & $cert_3$ & District3 & camera & 2012-06-12 07:22:45\\
12 & 2 & no & $cert_4$ & District1 & camera & 2012-04-07 14:33:28\\
$\cdots$ & $\cdots$ & $\cdots$ & $\cdots$ & $\cdots$ & $\cdots$ & $\cdots$\\
\hline
\end{tabular}
\end{center}
}

Furthermore, the data should be accessed by the legitimate users that can pose standard, select-from-where SQL queries like:
\begin{figure}%
\centering
\begin{minipage}{.4\columnwidth}%
\begin{center}
$Q1$
\end{center}
\begin{lstlisting}
select *
from T
where TypeId = 3
and Position = "District1";
\end{lstlisting}
\end{minipage}
\qquad
\begin{minipage}{.4\columnwidth}%
\begin{center}
$Q2$
\end{center}
\begin{lstlisting}
select ServiceId, TypeId
from T
where Position = "District1"
and Availability = "yes";
\end{lstlisting}
\end{minipage}%

\caption{Two conjunctive SFW queries}
\label{fig:queries}
\end{figure}

As an additional layer of security, the queries themselves are considered sensitive by the data owner $U$.
Points (i), (ii) (iii) are conflicting and to find a practical and efficient solution satisfying all of them is a challenging task.

In this work, we propose an application-level database encryption technique, based on an attribute-based cryptographic scheme, for processing obfuscated selection-projection
queries over an encrypted version of the table $T$, such that:
(i) data owner $U$ stores an encrypted version of the table $T$ on an non-private (possibly untrusted) RDBMS;
(ii) Each user $V$ entitled for running a query $Q$ over $T$ receives a corresponding decryption key $K_q$
encoding the query. That is,
only users satisfying a given access policy may access data satisfying a corresponding query, and \emph{only} such data.
(iii) The data owner issues the decryption keys in an initial setup phase;
(iv) The processing of the query is done by a trusted, separate query processor. Depending on architectural choices, such query processor may be
under the control of end users, or managed by a trusted third party, detached from the rdbms server;
(v) Data stored in the encrypted table may be modified by the ownwer $U$ without the need of re-encrypting the entire table, nor recomputing existing
decryption keys.
Note that in the proposed approach, access control is performed by the data owner by safely distributing to each user entitled to perform a query a corresponding decryption key. The encryption procedure is completely oblivious of the access control policy and this is a good feature since data can be encrypted even before potential readers with still to be defined access rights enroll into the system. Specifically, the addition of new readers does not force the data owner to modify the encryption of the current data.
Given that $U$ wants the encrypted data to be searched by queries like query $Q1$ and $Q2$ stated above, $U$ performs the
following operations:
it encrypts the tuples of table $T$ using an attribute-based encryption scheme that takes as additional input a vector $X$
containing (an encoding of) the values contained in the tuples. That is, there is a different vector $\x$ for every different row;
this has to be repeated as many times as there are attributes to be projected. In order to do this, we add a special attribute
whose domain ranges over the indices encoding the positions of attributes;
Afterwards, $U$ computes the corresponding secret restricted decryption keys, that depend on the values to be searched, encoded in a vector $Y$; 
since the attribute-based cryptographic scheme obfuscates the search values contained in the vector $Y$, it is referred to as
\emph{hidden vector encryption} (or \emph{HVE}) scheme.
The decryption keys are then distributed by $U$ (or some other trusted authority) to the users satisfying the corresponding access policy.
This is the last time in which users interact. After the decryption keys' distribution, the data owner can go offline.
Note that in the scenario just described the query are entirely defined by the data owner $V$, that fixes -- among other things -- the values to be 
searched in the table $T$.
However, in real-world database-backed applications, it is the case that the end user $V$ may specify the values to be searched filling empty
placeholders in an a-priori defined query. We show that a very simple modification of the proposed scheme allows the data owner $U$ to issue
a ``parametric'' token $\overline{K}$ corresponding to a query with unspecified search values, that are to be provided by the user receiving such token.
Note that the encryption depends on the data contained in the tuples and decryption keys only depend on the query. Note also that
it is possible to adjoin or delete tuples from the database without the need to re-encrypt its entire content, but only the modified content has to be
re-enciphered.
We remark also that a single encryption of the database is sufficient for answering different obfuscated queries.
The experimental results show that -- quite expectedly -- the space overhead for storing the encrypted data grows linearly with the number of
columns of the cleatext database. This linear overhead holds for the execution time of an encrypted query, as well.
The paper is structured as follows: in Section \ref{sec:rel} are discussed the most relevant related works; in Section \ref{sec:crypto}, a simplified
version of the attribute-based encryption scheme of Okamoto and Takashima suitable for our context is presented; in Section \ref{sec:privquery} we show how to use the above mentioned scheme for executing obfuscated SQL queries over an
encrypted database; in Section \ref{sec:archi} the main functional components of an architecture implementing our approach are presented; 
in Section \ref{sec:exp} the results of the experiments with our implementation are shown; finally, in Section \ref{sec:conc} the conclusions
are drawn, along with a description of future extensions we are already working on.

\section{Related works}
\label{sec:rel}
The ever-growing adoption of cloud-based data storage and management service has prompted 
several research efforts aiming at assuring an high security level of the stored data
using cryptographic mechanisms that do not assume the service to be trusted. 
The problem of how to provide an adequate security level to the stored data by deploying
cryptographic techniques has been addressed by the database research community as well as
the cryptographic research community, each of them focusing on different aspects that are
relevant for the respective community.
See, for example, the scenarios described in \cite{DBLP:conf/ccs/SamaratiV10}. 

There are several works in the database literature that deal with encrypted databases.
Seminal works like \cite{DBLP:journals/tods/DavidaWK81} and \cite{DBLP:journals/tods/BayerM76}
have underlined the need to devise novel mechanisms that complement the usual access control approaches
in order to achieve a greater level of security, by storing an encrypted version of the data, 
as well as the need to devise efficient indexing techniques that allow reasonable query performance over encrypted data.
The advent of fast networks and cheap online storage has made viable the management of 
encrypted data at application-level. One of the first works to present the paradigm \emph{database-as-a-service} is
\cite{DBLP:conf/sigmod/HacigumusILM02}. 
In \cite{DBLP:journals/pvldb/BajajS11} a database architecture based on a trusted hardware cryptographic module is presented.
In \cite{DBLP:conf/sosp/PopaRZB11}, the authors propose a full-fledged system for performing sql queries over an encrypted database.
Compared to such works, our approach is far less complex -- albeit with admittedly less search functionalitiesc-- while 
offering a strong degree of security. 
All the major commercial RDBMS releases provide functionalities to encrypt the data they store, see
for example \cite{OracleSecurityWhitePaper}. All commercial solutions are based on database-level encryption,
thus limiting the functionalities over encrypted data, that have to be deciphered at server side in order to
processing queries over them.
Apart from encryption techniques to deal with large quantities of data that are to be managed and searched over, issues
like key management and indexes over encrypted data have been addressed
In \cite{DBLP:journals/sigmod/ShmueliVEG09}, \cite{DBLP:reference/crypt/BouganimG11}
are presented two short surveys of the major challenges that are relevant to the design of 
encrypted database as well as useful reference architectures. 
In the cryptographic (and more broadly, security) research community, the problem of how to query an
an encrypted database has been viewed as a (very relevant) example of the broader problem
of computing over encrypted data. As it is well known, the first general, fully homomorphic encryption scheme
has been defined in \cite{DBLP:conf/stoc/Gentry09}. However, it is too impractical to be applied to real-world scenarios and a very
active research area in cryptography is to find 
Therefore, more specialized techniques have been proposed in order to solve the (less general) problem of searching over encrypted data.
One of the first works addressing the problem is \cite{DBLP:conf/eurocrypt/BonehCOP04}, in which the authors define a 
public key scheme that -- given a search keyword -- allows for the creation of a corresponding decryption key that tests whether
such keyword is contained in the ciphertext.
A subsequent work has introduced the notion of hidden vector encryption, in which it is possible to pose conjunctive and range queries
over encrypted data in suh a way that the query is itself obfuscated
\cite{DBLP:conf/tcc/BonehW07}.
Among the many works based on homomorphic encryption, we mention \cite{DBLP:conf/acns/BonehGHWW13}, that uses a
``limited'' (yet more efficient) version of homomorphic encryption to query an encrypted repository.
We point out though that homomorphic encyrption is at the moment very inefficient and thus the approaches based on it cannot be considered practical.
A similar approach to ours is described in
\cite{DBLP:journals/iacr/ZhengXA13}, where the authors propose an attribute-based encryption scheme for keyword searching over unstructured
textual data, along with a mechanism for proving the truthfulness of the search results.
An extensive amount of work has been done concerning how to access data without
disclosing sensitive information using anonymous credentials
see for example \cite{DBLP:conf/spw/BangerterCL04} and \cite{DBLP:conf/idman/CamenischDLNPP13}.
Finally, a related area is \textit{private information retrieval} in which the goal is to
preserve the privacy of queries. That is retrieve information from a database (typically represented as
an unstructured sequence of bits) without letting the database know anything about the query \cite{DBLP:journals/eatcs/Gasarch04}.
While it is a very interesting approach from a theoretical point of view, its practical applicability is rather scarce and it only allows very simple
queries (like retrieving a cell in a table) and does not 
enforce any control on what is accessed.

The approach presented in this work takes inspiration from attribute-based encryption schemes as defined, for example, in 
\cite{DBLP:conf/pairing/IovinoP08} or \cite{DBLP:conf/tcc/BonehW07}. However, since such schemes are optimized for a binary alphabet and
thus not very efficient over lager alphabets, such as those found in practice.
The techincal core of our contribution is a novel encryption scheme for 
HVE~\cite{DBLP:journals/joc/KatzSW13} that is based on the 
dual pairing vector space abstraction of \cite{OkTa} (which supports
very large alphabets) and allows for an
efficient amortized way of encrypting the rows of a table in such a way
that selection-projection queries on encyrpted data can be easily performed.
A self-contained presentation of the cryptographic tool we develop
is found in Section \ref{sec:crypto}.

\section{The cryptographic scheme constructions}
\label{sec:crypto}
In this section we describe our construction of HVE.
We start by describing the {\em dual pairing vector space} framework 
(the DPVS framework) of \cite{OkTa}.

\subsection{Dual Pairing Vector Space}
In the DPVS framework, 
we have an additive group $(\G,0)$  and a multiplicative group $(\G_T,1_T)$ of the 
same prime size $q$ 
and a bilinear map $\e:\G\times\G\rightarrow\G_T$.
That is, for $a,b,c\in\G$, we have
$$\e(a+b,c)=\e(a,c)\cdot\e(b,c)\ \text{ and } 
  \e(a,b+c)=\e(a,b)\cdot\e(a,c)$$
and 
$$\e(0,a)=1_T\ \text{ and } \e(a,0)=1_T.$$
The above imply that for all $s,t\in\F_q$ we have
$$\e(s\cdot a,t\cdot b)=\e(a,b)^{st}.$$

The bilinear map is extended to vectors over $\G$ as follows.
For two vectors $X=(x_1,\ldots,x_n)$ and $Y=(y_1,\ldots,y_n)$ over $\G$, define 
$$\e(X,Y)=\prod_{i=1}^n \e(x_i,y_i).$$
Note the abuse of notation by which we use $\e$ to denote the bilinear map defined over $\G$ and over
the vector space over $\G$.
Also, we observe that the extended bilinear map $\e$ is still bilinear in the sense that 
$$\e(X+Y,Z)=\e(X,Z)\cdot\e(Y,Z)\ \text{ and } \e(X,Z+Y)=\e(X,Z)\cdot\e(X,Y)$$
for all vectors $X,Y,Z$.

For a fixed $g\in\G$, 
we define  the {\em canonical base} 
$A_1,\ldots,A_n$ 
with respect to $g$ where, for $i=1,\ldots, n$,
$$A_i=(\underbrace{0,\ldots,0}_{i-1},g,\underbrace{0,\ldots,0}_{n-i}).$$
Notice that 
$$\e(A_i,A_i)=\e(g,g)\ \text{ and } i\ne j \Rightarrow \e(A_i,A_j)=1_T.$$

Let $B$ and $B^\star$ be 
two $n\times n$ matrices with columns $B=(B_1,\ldots,B_n)$ and $B^*=(B_1^\star,\ldots,B_n^\star)$ and
let $\psi\in\F_q$. We say that $(B,B^\star)$ is a pair of {\em $\psi$-orthogonal matrices} if, 
for all $1\leq i<j\leq n$, 
$$\e(B_i,B_i^\star)=g_T^\psi\ \text{ and } \e(B_i,B_j^\star)=1_T, $$
where we set $g_T=\e(g,g)$.
A pair of $\psi$-orthogonal matrices can be constructed as follows.
Let $X=(x_{i,j}),X^\star=(x_{i,j}^\star)\in\F_q^{n\times n}$ be matrices such that 
$$X^T\cdot X^\star=\psi\cdot I.$$
for $i=1,\ldots,n$, vectors $B_i$ and $B_i^\star$ are defined as follows
$$B_i=\sum_{j=1}^n x_{i,j} A_j\text{ and } 
  B_i^\star=\sum_{j=1}^n x_{i,j}^\star A_j.$$

For a vector $(x_1,\ldots,x_n)\in F_q$ and a matrix $B=(B_1,\ldots,B_n)$, we define the vector
$$(x_1,\ldots,x_n)_B=\sum_{i=1}^n x_i B_i.$$

\subsection{HVE in the DPVS framework}
\label{sec:scheme}
In this section we first define the concept of an HVE encryption scheme and 
then give an implementation using the DPVS framework.

The HVE function over $\F_q^\ell$ is defined as follows:
for $X=(x_1,\ldots,x_\ell)\in\F_q^\ell$ and 
$Y=(y_1,\ldots,y_\ell)\in(\F_q\cup\{\star\})^\ell$, we define 
$$\HVE(X,Y)=1\ \text{ iff } \forall i\ y_i\ne\star\Rightarrow x_i=y_i.$$

\begin{definition}
An HVE encryption scheme is a quadruple of algorithms 
$\HVE=($\textsf{Setup},\textsf{Encrypt}, \textsf{KeyGen}, \textsf{Decrypt}$)$ with the following 
syntax:

\begin{enumerate}
\item The \textsf{Setup} algorithm takes integers $q$ and $\ell$ and returns 
a {\em master public key} $\mathsf{MPK}$ and 
a {\em master secret key} $\mathsf{MSK}$.
\item The \textsf{Encrypt} algorithm takes a {\em plaintext} $M$, 
an {\em attribute} vector $X=(x_1,\ldots,x_\ell)\in\F_q^\ell$ and
a {master public key} $\mathsf{MPK}$ and returns a {\em ciphertext} $\mathsf{Ct}$.
\item The \textsf{KeyGen} algorithm takes an {\em attribute} vector
$Y=(y_1,\ldots,y_\ell)\in(\F_q\cup\{\star\})^\ell$ and  master key $\mathsf{MSK}$
returns a {\em key} $K$ for $Y$.
\item The \textsf{Decrypt} algorithm takes a {ciphertext} $\mathsf{Ct}$ for 
plaintext $m$ and attribute vector $X$ computed using master public key $\mathsf{MPK}$
and a key $K$ for attribute vector $Y$ computed using master secret key $\mathsf{MSK}$
and returns the value $m$ iff $\HVE(X,Y)=1$.
\end{enumerate}
\end{definition}

We are now ready to describe our implementation $\HVE$ of an HVE encryption scheme.
We assume that a DPSV framework $(\G,\G_T,\e,q)$ is given.

\begin{enumerate}
\item \textsf{Setup}$(\ell)$. 
Randomly choose $\psi\in\F_q$ and $g\in\G$ and set $g_T=\e(g,g)^\psi$.
For $i=0,\ldots,\ell$ generate a pair $(B^i,C^i)$
of $\psi$-orthogonal $3\times 3$ matrices.
Return $\mathsf{MPK}=(B_0,\ldots,B^\ell,g_T)$ and 
    $\mathsf{MSK}=(C^0,\ldots,C^\ell)$.

\item \textsf{Encrypt}$(m,X,\mathsf{MPK})$. 
We assume $m\in\G_T$ and $X=(x_1,\ldots,x_\ell)\in\F_q^\ell$.

Randomly choose $z,w_0,\ldots,w_\ell\in\F_q$, set 
$$c=g_T^z\cdot m\ \text{\rm and } c_0=(w_0,z,0)_{B^0}$$
and, for $t=1,\ldots,\ell,$ set 
$$c_t=(w_t,w_t\cdot x_t,w_0)_{B^t}.$$
Return $\mathsf{Ct}=(c,c_0,c_1,\ldots,c_\ell)$.

\item \textsf{KeyGen}$(Y,\mathsf{MSK})$.
Assume $Y=(y_1,\ldots,y_\ell)\in(\F_q\cup\{\star\})^\ell$ and 
let $S$ be the set of $1\leq t\leq \ell$ such that $y_t\ne\star$.

Pick random $\eta\in\F_q$ and, for $t\in S$, 
pick random $d_t,s_t\in\F_q$ and set $s_0=-\sum_{t\in S} s_t$. 
Set 
$$k_0=(s_0,1,\eta)_{C^0}\ \text{ and } k_t=(d_t\cdot y_t,-d_t,s_t)_{C^t}.$$
Return $K=(k_0,(k_t)_{t\in S})$.

\item \textsf{Decrypt}$(K,\mathsf{Ct})$. 
Write $K$ as $K=(k_0,(k_t)_{t\in S})$ and 
$\mathsf{Ct}$ as $\mathsf{Ct}=(c,c_0,\ldots,c_\ell)$.
Return
$$\frac{c}{\e(k_0,c_0)\cdot\prod_{i\in S} \e(k_i,c_i)}.$$
\end{enumerate}

Let us now show that our scheme is correct.

Suppose that $\HVE(X,Y)=1$.  Then 
by the $\psi$-orthogonality of $B^0$ and $C^0$ we have
$$\e(k_0,c_0)=g_T^{s_0\cdot w_0+z}.$$
Moreover, the $\psi$-orthogonality of $B^i$ and $c^i$ gives
$$\e(k_t,c_t)=g_T^{d_t\cdot w_t\cdot(x_t-y_t)+w_0\cdot s_t}.$$
Therefore if $x_t=y_t$ for $t\in S$ we have 
$$\prod_{t\in S} \e(k_t,c_t)=g_T^{w_0\cdot \sum_{t\in S} s_t}=g_T^{-s_0\cdot w_0}$$
which implies that the \textsf{Decrypt} algorithm returns $m$.
On the other hand if $x_t\ne y_t$ for some $t\in S$ the 
\textsf{Decrypt} algorithm returns a random value in $\G_T$.

For security we observe that a ciphertext for plaintext $m$ with attribute vector $X$
does not reveal any information on $m$ and on attribute vector $X$. 
On the other hand, no security guarantee is made for a key $K$.

\subsection{An amortized scheme}
\label{sec:amort}
In our construction of secure database queries we will often have to encrypt 
$n$ messages $m^1,\ldots,m^n\in\G_T$ with closely related attributes. 
More specifically,  
message $m^j$ is encrypted with attributes
$(x_1,\ldots,x_\ell,x^j_{\ell+1})\in\F_q^{\ell+1}$; that is, 
the attributes of two messages coincide except for the $(\ell+1)$-st.
If we use the HVE implementation of the previous section, 
the sum of the sizes of the ciphertexts of all messages is $\Theta(\ell\cdot n)$. 
In this section we describe a scheme that reduces the size to $\Theta(\ell+n)$.

\begin{enumerate}
\item \textsf{Setup}$^{\mathsf{am}}$. 
Same as $\mathsf{Setup}$.

\item \textsf{Encr}$^{\mathsf{am}}
(m^1,\ldots,m^n\in\G_T,x_1,\ldots,x_\ell,x^1_{\ell+1},\ldots,x^n_{\ell+1},{\mathsf{MPK}})$.

Randomly choose $z,w_0,\ldots,w_\ell\in\F_q$ and set 

$$c_0=(w_0,z,0)_{B^0}\mbox{\rm \quad and\quad } c_t=(w_t,w_t\cdot x_t,w_0)_{B^t}
\mbox{\rm\quad for\quad} t=1,\ldots,\ell.$$

The encryption of the $j$-th message $m^j$ is computed as follows:
Pick random $z^j,w^j,w_0^j\in\F_q$ and set 
$$c^j=g_T^{z+z^j}\cdot m^j$$
and 
$$c_0^j=(w_0^j,z^j,0)_{B^{0,\star}}\mbox{\rm\quad and\quad } 
  c_{\ell+1}^j=(w^j,w^j\cdot x_{\ell+1}^j,w_0^j)_{B^{\ell+1}}.$$
The cumulative ciphertext consists of 
$$\mathsf{Ct}=(c_0,c_1,\ldots,c_\ell,(c^1,c_0^1,c_{\ell+1}^1),\ldots,
                         (c^n,c_0^n,c_{\ell+1}^n)).$$
The ciphertext corresponding to $m^j$ is 
$$\mathsf{Ct}^j=(c_0,c_1,\ldots,c_\ell,(c^j,c_0^j,c_{\ell+1}^j)).$$

\item \textsf{KeyGen}$^{\mathsf{am}}(Y,\mathsf{MSK})$.

Write $Y$ as $Y=(y_1,\ldots,y_\ell,y_{\ell+1})$ and  
let $S$ be the set of $1\leq t\leq\ell$ such that $y_t\ne\star$.
We assume that $y_{\ell+1}\ne\star$ and we stress that $\ell+1\not\in S$.

Randomly choose $\eta\in\F_q$ and, for $t\in S$, randomly choose 
$d_t,s_t\in\F_q$ and set $s_0=-\sum_{t\in S} s_t$. Set
$$k_0=(s_0,1,\eta)_{C^0}\quad\mbox{\rm{and}}\quad 
        k_t=(d_t\cdot y_t,-d_t,s_t)_{C^t}.$$
Randomly choose $s_0^\star,\eta^\star,d_{\ell+1}\in\F_q$ and output
$$k_0^\star=(s_0^\star,1,\eta^\star)_{C^{0,\star}} 
    \quad\mbox{\rm{and}}\quad 
    k_{\ell+1}=(d_{\ell+1}\cdot y_{\ell+1},-d_{\ell+1},-s_0^\star)_{C^{\ell+1}}.$$

Return key $K$ 
$$K=(k_0,(k_t)_{t\in S},k_0^\star,k_{\ell+1}).$$

\item \textsf{Decrypt}$^{\mathsf{am}}(K,\mathsf{Ct})$.
Write $K=(k_0,(k_t)_{t\in S},k_0^\star,k_{\ell+1})$ 
and $\mathsf{Ct}=(c_0,c_1,\ldots,c_\ell,(c^j,c_0^j,c_{\ell+1}^j)_{j=1}^n)$.
Compute $m^j$ as 
$$\frac{c^j}{
\e(c_0^j,k_0^\star)\cdot
\e(c_{\ell+1}^j,k_{\ell+1})\cdot
\e(c_0,k_0)\cdot\prod_{t\in S}\e(c_t,k_t)}.$$
\end{enumerate}

For correctness, suppose $\HVE(X,Y)=1$.
By the $\psi$-orthogonality we have
$$\e(c_0,k_0)=g_T^{s_0\cdot w_0+z}.$$
and, for $t\in S$
$$\e(c_t,k_t)=g_T^{s_t\cdot w_0}.$$
Therefore 
$$\prod_{t\in S} \e(c_t,k_t)=g_T^{\sum_{t\in S}s_t\cdot w_0}=g_T^{-s_0\cdot w_0}.$$
and thus 
$$\e(c_0,k_0)\cdot \prod_{t\in S}\e(c_t,k_t)=g_T^z.$$
Moreover we have 
$$
\e(c_0^j,k_0^\star)=g_T^{s_0^\star\cdot w_0^j+z^j}
\quad{\mbox{\rm and}}\quad
\e(c_{\ell+1}^j,k_{\ell+1})=g_T^{-s_0^\star\cdot w_0^j}
$$ 
and thus
$$
\e(c_0^j,k_0^\star)\cdot
\e(c_{\ell+1}^j,k_{\ell+1})=g_T^{z^j}.$$

\section{Private queries}
\label{sec:privquery}
In this section we describe how a table is encrypted using the HVE scheme 
of Section~\ref{sec:crypto}.
Let us start by fixing our notation.

We assume that data owner $U$ holds table $T$, 
composed of $l$ columns $A_1,\ldots,A_l$ and of $u$ rows $R^1,\ldots,R^u$
and we write the $i$-th row $R^i$ as 
$R^i=\langle v_i^1,\ldots,v^i_l\rangle$.

\paragraph{Encrypting a table.}
We assume that the data owner $U$ has selected a DPVS framework with groups of 
size $q$ and an authenticated private-key block cipher 
$\mathsf{ABC}=(E,D)$ with key length $k<q$.

\begin{enumerate}
\item Generating the system parameters.

$U$ runs algorithm $\mathsf{Setup}^{\mathsf{am}}(q,l)$ and obtains a pair of 
master public and secret key $(\mathsf{MPK},\mathsf{MSK})$.

\item Encrypting row $R=\langle v_1,\ldots,v_l\rangle$.

$U$ picks $l$ random keys $k_1,\ldots,k_l$ for $\mathsf{ABC}$ and 
use them to encrypt $R$. Specifically, 
compute $\tilde R=\langle \tilde v_1,\ldots,\tilde v_l\rangle$, 
where $\tilde v_i=E(k_i,v_i)$ for $i=1,\ldots,l$.

Then $U$ encrypts keys $k_i$ using the HVE scheme. Specifically, 
for $i=1,\ldots,l$, $U$ computes $\tilde k$ by setting
$$\tilde k=\mathsf{Encrypt^{am}}(k_1,\ldots,k_l,v_1,\ldots,v_l,1,\ldots,l,
\mathsf{MPK}).$$
That is, key $k_i$ is encrypted with attribute vector 
$(v_1,\ldots,v_l,i)$.

The encryption of row $R$ consists of the pair $(\tilde R,\tilde k)$ and 
the encrypted table is simply the sequence 
$\tilde T=
\langle(\tilde R^1,\tilde k^1),\ldots,(\tilde R^u,\tilde k^u)\rangle$ of
the encrypted rows.

\end{enumerate}

\paragraph{Generating a key for a  selection-projection query.}
A typical selection-projection query (like $Q2$ in Figure \ref{fig:queries}
``\textbf{select} ServiceId, TypeId \textbf{from} T \textbf{where} Position = `District1' \textbf{and} Availability = `yes' '')
 is described by specifying the search values, along with
 ``don't care'' entries $\star$, and the columns that are to be projected in the corresponding attribute vectors
(one for every column to be projected).

More specifically, consider a query that specifies value 
$a_i$ for $i=1,\ldots,l$ 
(it is possible that $a_i=\star$ for some values $i$ corresponding to 
don't care entries in the query) and column $c$ to be projected. 
We assume without loss of generality that only one columns is 
to be projected; in general 
one computes a different key for each column to be projected.
For example, a selection-only query, like query $Q_1$:``\textbf{select} * \textbf{from} T \textbf{where} TypeId = 3 \textbf{and} Position = `District 1' '',
is a special case of selection-projection query in which \emph{all} the columns of the table are to be projected. The key for such a query is obtained
by running the procedure described below with $c=1,\ldots,l$.

The data owner $U$ releases key $K$ for the query $Q1$
by running ${\mathsf{KeyGen^{am}}}(Y,\mathsf{MSK})$ for 
attribute vector $Y=(a_1,\ldots,a_l,c)$ and using master secret key
$\mathsf{MSK}$ computed as part of the generation of the system 
parameter generation. The key $K$ is then sent to an user $V$ entitled  
to execute the query.

We point out that in the description above we have implictly assumed that
values $a_i$ are elements of the field $\F_q$ or $\star$. 
This does not hold in general as in 
most applications values $a_i$ are strings over an alphabet; thus we
derive values $a_i$ by applying an hash function that for each string 
returns an element of $\F_q$.

\paragraph{Executing a selection-projection query.}
Upon receiving key $K$, the user $V$ applies $K$ to the encrypted
table as follows.
For $i=1,\ldots,u$, row $i$ $(\tilde R^i,\tilde k^i)$ of the encrypted table 
is used to compute  
$\hat k_c^i$ by selecting the $c$-th output of the 
${\mathsf{Decrypt^{am}}}$ algorithm on input $K$ and $\tilde k^i$. 
Key $\hat k_c^i$ is then used to run the decryption algorithm $D$ for 
{\sf{ABC}} to decrypt $\hat v^i_c$.
If the decryption algorithm succeeds then it returns $v^i_c$. 
Otherwise the row is not selected.

Indeed, if the key attribute vector of $K$ matches  
the attribute vector used to obtain $\tilde k^i$ 
then this means that row $i$ is to be selected and 
thus $\hat k_c^i=k_c^i$ and the decryption algorithm $\mathsf{Decrypt^{am}}$ does not fail.
If instead the key attribute vector of the key $K$ does not match 
the attribute vector of the row (that is, the row is not to be selected)
then $\hat k_c^i$ is a random element of $\G_T$ and thus with 
very high probability the decryption algorithm fails.

\subsection{Parametric queries}
\label{sec:paraquery}
In the previous sections, we have assumed that the
data owner, in constructing the key $K$, 
completely determines the query that can be run over the encrypted database.
That is, the values that are to be searched 
(and their corresponding columns) and the columns that are to be returned. 
However, in real-world, database-backed applications what usually happens 
is that the user asking for the query execution has the ability to
fix by itself the values to be searched.
Specifically, the data owner might want to be able to generate a key that
allows to search a specified column (or set of columns) for values to 
be specified later.
We remark that the way tokens are computed by the $\mathsf{KeyGen}$ 
procedure easily allows for a two-step computation of a token 
encoding a given query: 
namely, the first step is performed by the data owner which computes 
an intermediate, ``{\em parametric}'' decryption key; 
the second step then consists in specifying the parameters into the 
parametric decryption key thus obtaining a complete decryption key. 
More precisely, we observe that the computation of decryption key components $k_t$, for $1\leq t\leq l$ (the ones that depend on the values to be searched
(see Section \ref{sec:crypto}) can be performed by the following modified 
version of the \textsf{KeyGen}$^{\mathsf{am}}$:

\noindent\textsf{KeyGen}$^{\mathsf{am}}_\mathsf{{par}}$:
The decryption key is computed by a two-step process: (i) for every decryption key $K$, 
for $t\in S$, pick random $s_t, d_t$ from $\F_q$ and set $\overline{k_t}=(d_t, -d_t, s_t)_{C^t}$
($k_0$ is defined as in  \textsf{KeyGen}$^{\mathsf{am}}$); (ii) multiply element-wise
the vector $\overline{k_t}$ (which is composed by three elements) with the vector $(y_t, 1, 1)$, obtaining the component $k_t$ of the decryption
key $K$.

We note that Steps (i) and (ii) may be performed by different entities. In fact,
with respect to Query $Q2$, following Step 1 user $U$ computes 
the parametric decryption keys $\overline{K_1}$, $\overline{K_2}$ that correspond to the parametric query 
``select columns \textit{ServiceId} and \textit{TypeId} of all the rows of the table $T$ such that columns \textit{Availability}
and \textit{Position} contain values `$value1$' and `$value2$' respectively''.
A  (possibly different) user $V$ then 
after having received $\overline{K_1}$, $\overline{K_2}$ from $U$, computes the decryption keys $K_1$, $K_2$ by specifying values $y_1$, $y_2$ corresponding to
search values  `\textit{yes}', `\textit{District3}'.
In other words the data owner $U$ can delegate restricted 
search capabilities to user $V$.

For the security we observe that the parametric key does not allow
to generate keys other than the ones that can be obtained by specifying the
values of the parameters.



\section{Architectural Overview}
\label{sec:archi}
\begin{figure}[htb]
\centering
  \includegraphics[width=.5\columnwidth]{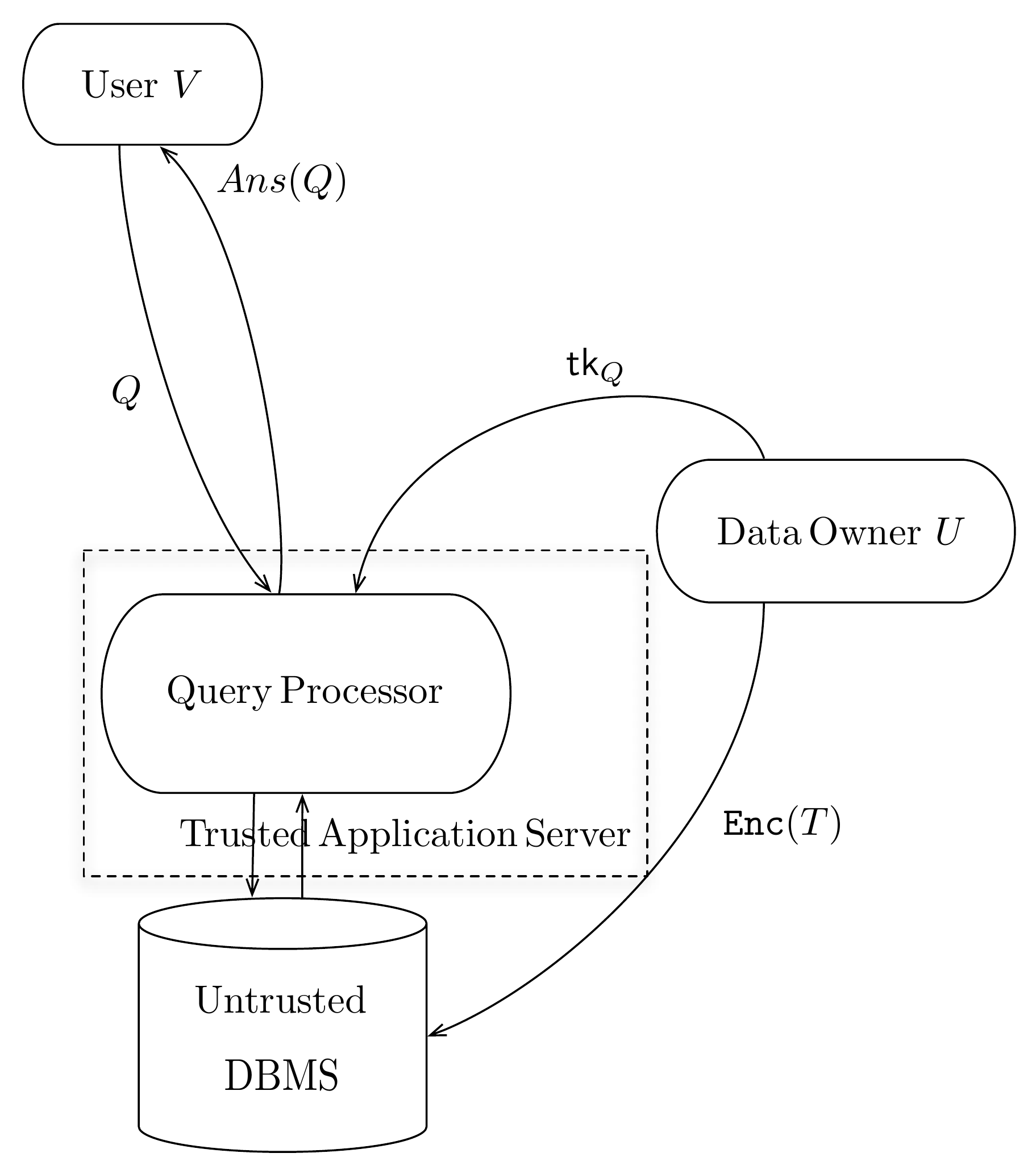}
\caption{The high-level architecture}
  \label{fig:archi}
\end{figure}

The high-level overview of the architecture (as shown in Figure \ref{fig:archi}) assumes that there are users $V$, $V',\ldots$, a data owner $U$,
and a server split in two components:
a query processor $P$ hosted on a trusted, separate application server, and a (possibly untrusted) DBMS server $S$.
The encryption and decryption operations are thus delegated outside the DBMS server (compare with, for example, with the application-level architecture in the taxonomy presented in \cite{DBLP:reference/crypt/BouganimG11}).
For sake of simplicity, we assume that the data owner $U$ generates the 
public parameters, as well as the key material deployed in the execution of the cryptographic schemes described in the previous sections
\footnote{As a more realistic setting, we may assume that the tokens are generated by users having proper credentials or by other trusted third parties
that check such credentials}.

Also, we assume that the proxy query processor $P$ (whose task is to decrypt the (parts of the) rows having matching values with
the encryption attribute vector) communicates with the other components via secure and authenticated channels. 
Confidentiality of both data and queries hold as long as the proxy $P$ and the server $S$ do not collude.
We now illustrate the workflow occurring among the data owner, the trusted proxy query processor, the untrusted server and users who wish to pose queries to the encrypted database.
We consider the non-amortized case, being the amortized one rather similar in the sequence of actions to be performed.

\begin{figure}[htb]
  \centering
  \includegraphics[width=.5\columnwidth]{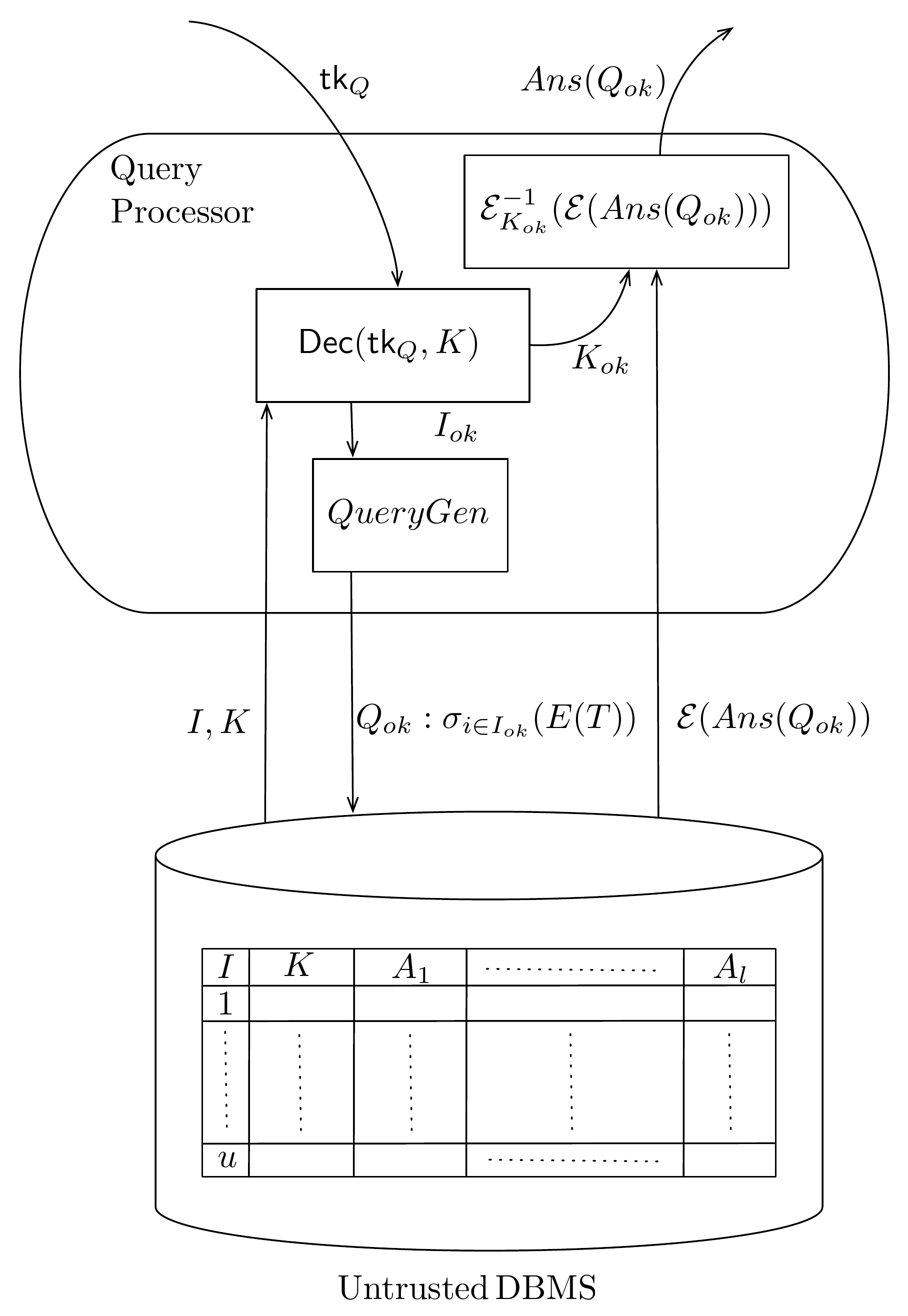}
  \caption{The query processor structure}
  \label{fig:qp}
\end{figure}

The \emph{data owner} $U$ runs the \textsf{Setup} procedure (see Section \ref{sec:crypto}), producing the secret master key $C^0,\ldots,C^l$ and the
public key $g_T=\e(g,g)^\psi,B^0,\ldots,B^l$, $u$ randomly generates the secret \textsf{ABC} keys $k_1,\ldots,k_u$, as well . $U$ then encrypts the table in two steps: (i) rows are 
encrypted with symmetric scheme \textsf{ABC} and (ii) the corresponding secret keys are encrypted with the public keys using the HVE-based \textsf{Encrypt} procedure.
The encrypted database $\textsf{Enc}(T)$ (composed of the \textsf{ABC}-encrypted rows and the HVE-encrypted \textsf{ABC} keys and an additional column that stores a row counter)
is uploaded in the untrusted DBMS server. Afterwards, given a query
$Q$, $U$ proceeds in computing the corresponding token $\textsf{tk}_Q$, using the secret master key and the the vector containing the values specified in the query $Q$).
Finally, the decryption key $K_Q$ is sent to the query processor. At this point, the data owner $u$ may go offline.
Upon receiving a request  from user $V$ for executing query $Q$, The \emph{proxy} query processor $P$ performs the following steps
(see Figure \ref{fig:qp}):

\begin{enumerate}
\item[(i)] retrieve with a table scan from the encrypted database (stored in the servers $S$) the columns containing the row
counter and the \textsf{HVE}-encrypted \textsf{ABC} keys (which we denote $I$ and $KHVE$, respectively) from the encrypted database;
\item[(ii)] execute the procedure \textsf{Decrypt} with $K_Q$ on each \textsf{HVE}-encrypted \textsf{ABC} key in $KHVE$; for every successfully decrypted AES key,
put the corresponding value of the row counter into $I_{ok}$; put the deciphered
\textsf{ABC} keys in $KHVE_{ok}$;
\item[(iii)] execute the following SQL query over the encrypted database:

\[
Q_{ok}=\textbf{select\,}\ast\,\textbf{from\,}E(T)\,\textbf{where\,}I\,\textbf{in\,}I_{ok}
\]

\item[(iv)] retrieve the answer $Ans(Q_{ok})$;
\item[(v)] decipher the \textsf{ABC}-encrypted rows in $Ans(Q_{ok})$ with the keys in $KHVE_{ok}$;
\item[(vi)] send the deciphered rows to the issuer of query $Q$ 
\end{enumerate}

\paragraph{The case of parametric queries.} Note that the workflow presented in the previous section requires that the query $Q$ to be fully 
specified by the data owner. In particular, $U$ 
knows the value(s) to be searched. 
In reality, what happens -- as already explained in Section \ref{sec:paraquery} -- is that user $V$ specifies the search values in a query $Q$,
provided by the data owner $U$. 
The architecture just presented allows as well for the execution of a query following the steps specified in Section \ref{sec:paraquery}: 
the data owner $U$ sends the parametric token $\bar{\textsf{tk}_Q}$ to the proxy $P$. When user $V$ wants to execute the query $Q$, it sends (along with
s proper request) the vector $Y$ (containing the values to be searched) to $P$, that uses it to form the
decryption key $K_Q$. 
%

\section{Experimental evaluation}
\label{sec:exp}
We now describe the experimental results of the application of the schema presented in Section~\ref{sec:crypto} and \ref{sec:privquery}.
We implemented the schemes
presented in Section \ref{sec:crypto} in Python, using the Charm library\footnote{\url{http://www.charm-crypto.com/Main.html}}, using the
relational database SQLite\footnote{\url{http://www.sqlite.org}}.
Charm is a Python-based library that offers support for the rapid prototyping
and experimentation of cryptographic schemes. In particular, we rely on the support
offered by Charm for dealing with pairings, which is in turn based on the
well-known C library PBC\footnote{\url{http://crypto.stanford.edu/pbc/}}.
The implementation has been tested on a machine with a
2-core 2GHz Intel Core i7 processor with 8 GB 1600 MHz DDR3 RAM, running OS X 10.9.1.

\paragraph{Asymptotic complexity.}
Tables \ref{tab:amo} and \ref{tab:nonamo} show the complexities of the procedures presented respectively in Section \ref{sec:scheme} and Section \ref{sec:amort}. 
The operations we take into account in measuring time complexity are: pairings (denoted as $P$),
exponentiations in $G_T$ (denoted as $EG_T$), row products (denoted $RW$, see Point 6 in Section \ref{sec:crypto}).
As for space complexity, we denote with $|g|$ and $|g_T|$ the size of elements in $G$ and $G_T$. Given a SWF query to be privately executed over an encrypted version of table
$T$, 
we remind that $c$, $t$ anc $l$ respectively denote the number of projected columns in the \textbf{select} clause, the number of columns in the table $T$, specified in the \textbf{from} clause and 
the number of search predicates in the \textbf{where} clause. 

\begin{table}
\centering
\begin{tabular}{|c||c|c|}
\hline
& number of operations & output size \\ \hline\hline
\textsf{Setup} & $1\cdot P$ & $|g_T|+18l\cdot|g|$\\ \hline
\textsf{Enc} & $EG_T+(l+1)\cdot RW$ & $|g_T|+10l\cdot|g|$\\ \hline
\textsf{KeyGen} & $(t+1)\cdot RW$ & $3(t+1)\cdot |g|$\\ \hline
\textsf{Dec} & $3(t+1)\cdot P$ & \\
\hline
\end{tabular}
\vspace{10pt}
\caption{Asymptotic complexity of non-amortized scheme}
\label{tab:amo}
\end{table}

\begin{table}
\centering
\begin{tabular}{|c||c|c|}
\hline
& number of operations & output size \\ \hline\hline
\textsf{Setup} & $1\cdot P$ & $|g_T|+18l\cdot|g|$\\ \hline
\textsf{Enc} & $l\cdot EG_T+(l^2+l)\cdot RW$ & $|g_T|+13l\cdot|g|$\\ 
\hline
\textsf{KeyGen} & $c(t+1)\cdot RW$ & $3(t+3)\cdot |g|$\\ \hline
\textsf{Dec} & $3c(t+1)\cdot P$ & \\
\hline
\end{tabular}
\vspace{10pt}
\caption{Asymptotic complexity of amortized scheme}
\label{tab:nonamo}
\end{table}

\begin{figure}[htb]
  \centering
  \includegraphics[width=.8\columnwidth]{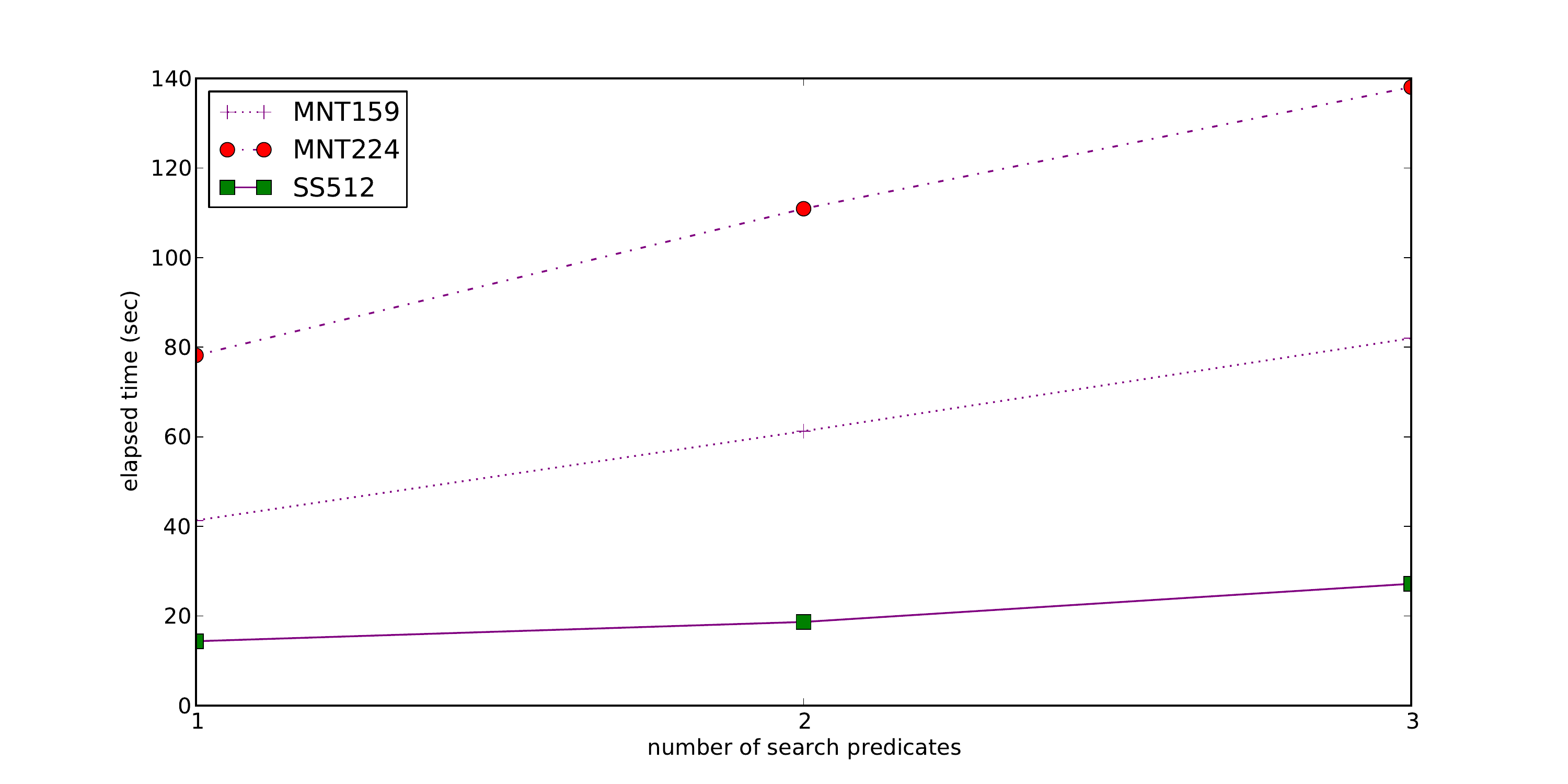}
  \caption{Queries execution time}
  \label{fig:time}
\end{figure}

\paragraph{Executions with real data.}
We have executed several tests on datasets of growing sizes, deploying different curve
parameters. More precisely, we have generated three synthetic relational tables of size
$234,554$, $463,999$ and $934,347$ bytes respectively; we have encrypted them at row level with
AES and subsequently we have encrypted the corresponding 256-bit AES keys with HVE-IP using as
curves parameters MNT159 ($G_1$, $G_2$\footnote{In this case, the pairing is asymmetric},
$G_T$ elements' bitsizes are respectively $159$, $477$, $954$), SS512
($G_1$, $G_2$, $G_T$ elements' bitsizes are respectively $512$, $512$, $1024$) and 
MNT224 (with $G_1$, $G_2$, $G_T$ elements' bitsizes of $224$, $672$, $1344$)
MNT159 parameters have a security level equivalent to 954-bit DLOG, while SS512 and MNT have security levels
equivalent respectively to 1024 and 1344-bit DLOG.
In Figure \ref{fig:time}, are shown the execution times of encrypted queries corresponding, respectively, to
the following SQL queries with search conditions of increasing length:

\noindent (i) \textbf{select * from} $T$ \textbf{where} ServiceId = 42;

\noindent (ii) \textbf{select * from} $T$ \textbf{where} ServiceId = 42 \textbf{and} TypeId = 3;

\noindent (iii) \textbf{select * from} $T$ \textbf{where} ServiceId = 42 \textbf{and} TypeId = 3 \textbf{and} Availability = `no';

\begin{figure}[htb]
  \centering
  \includegraphics[width=.8\columnwidth]{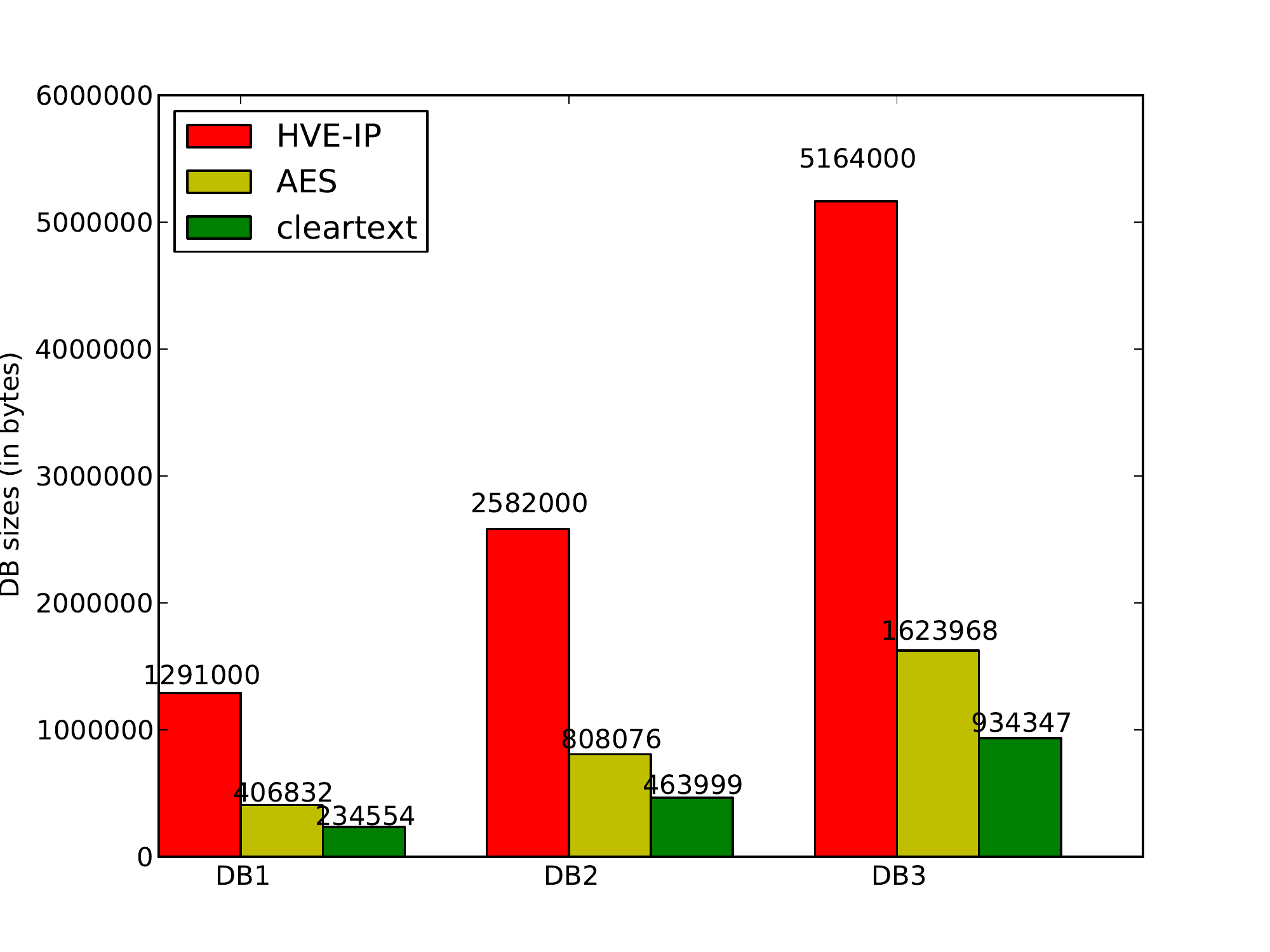}
  \caption{Encrypted database space overhead}
  \label{fig:space}
\end{figure}

As it is expected, for a given row in the cleartext table, the encryption execution time linearly depends on the number of 
table columns and the major time cost is due to the computation of pairings.
The queries have been performed on the largest database. 
We remind that an encrypted query execution is composed of the following steps (see Section 
\ref{sec:archi}): (i) retrieve from the encrypted database the rows index and the the hve-ip-encrypted aes keys,
(ii) for every retrieved encrypted key, run the decryption procedure with the attribute vector encoding the query and, for each key
that has a successful match, store the corresponding row index in the index result set $I_{ok}$, (iii) retrieve from the encrypted database the aes-
encrypted rows whose row indexes are contained in  $I_{ok}$, (iv) finally, decipher the rows with the corresponding aes keys that have been successfully 
decrypted in Step (iii).
In Figure \ref{fig:space}, the space overhead of the encrypted database is shown, resulting from the AES encryption of the
data stored in the plaintext database and the subsequent encryption of the corresponding AES keys using HVE-IP. Again as
expected, the expansion factor linearly depends on the number of columns employed by the HVE-IP encryption procedure.
With respect to the datasets used in the experiments, the expansion factor of the hve-ip encryption is roughly $5.5$.
Such factor is directly proportional on the number of columns in the cleartext database schema -- as already pointed out
in Table \ref{tab:nonamo} -- and inversely proportional to the size of the cleartext database. This is due, of course, to the fact
that values in the cleartext database (which can be of arbitrary size) are mapped to $\mathbb{F}_q$ elements (which are of fixed
size). As a rule of thumb for decreasing the size of the encrypted database, one 
should limit the columns on which the encryption depends only on the
ones that are actually involved in the search predicates.


\section{Conclusions}
\label{sec:conc}
In this work we have presented an attribute-based encryption scheme based on the DPVS framework and we have used it to define a system for
privately query with standard SQL conjunctive queries an encrypted, off-the-shelf, non-private relational
database without the need to
decipher the stored data during the execution of the intermediate query processing steps.
We have implemented our scheme using a standard relational RDBMS, and we provide experimental 
results that show the feasibility of our approach.
Regarding the extensions of the present work, 
we are currently finishing a full-fledged implementation in C in a client-server architecture, that will
provide us more realistic experimental results. 
also, we are currently working on privately executing (i) join queries
and (ii) aggregate queries on encrypted tables. Also, we are investigating ways for deploying 
indexes 
in order to avoid full table scans; also we plan to add mechanisms for verifying whether the query processor has 
faithfully executed the query.

\bibliographystyle{splncs}
\bibliography{hvedbbib}

\begin{thebibliography}{10}

\bibitem{DBLP:conf/ccs/SamaratiV10}
Samarati, P., di~Vimercati, S.D.C.:
\newblock Data protection in outsourcing scenarios: issues and directions.
\newblock In: Proceedings of the 5th ACM Symposium on Information, Computer and
  Communications Security, (ASIACCS), Beijing, China. (2010)  1--14

\bibitem{DBLP:journals/tods/DavidaWK81}
Davida, G.I., Wells, D.L., Kam, J.B.:
\newblock A database encryption system with subkeys.
\newblock ACM Trans. Database Syst. \textbf{6}(2) (1981)  312--328

\bibitem{DBLP:journals/tods/BayerM76}
Bayer, R., Metzger, J.K.:
\newblock On the encipherment of search trees and random access files.
\newblock ACM Trans. Database Syst. \textbf{1}(1) (1976)  37--52

\bibitem{DBLP:conf/sigmod/HacigumusILM02}
Hacig{\"u}m{\"u}s, H., Iyer, B.R., Li, C., Mehrotra, S.:
\newblock Executing sql over encrypted data in the database-service-provider
  model.
\newblock In: Proceedings of the ACM SIGMOD Conference on Management of Data.
  (2002)  216--227

\bibitem{DBLP:journals/pvldb/BajajS11}
Bajaj, S., Sion, R.:
\newblock Trusteddb: A trusted hardware based outsourced database engine.
\newblock PVLDB \textbf{4}(12) (2011)  1359--1362

\bibitem{DBLP:conf/sosp/PopaRZB11}
Popa, R.A., Redfield, C.M.S., Zeldovich, N., Balakrishnan, H.:
\newblock Cryptdb: protecting confidentiality with encrypted query processing.
\newblock In: Proceedings of the 23rd ACM Symposium on Operating Systems
  Principles (SOSP). (2011)  85--100

\bibitem{OracleSecurityWhitePaper}
Corp., O.:
\newblock Oracle advences security transparent data encryption best practices.
\newblock white paper (2012)

\bibitem{DBLP:journals/sigmod/ShmueliVEG09}
Shmueli, E., Vaisenberg, R., Elovici, Y., Glezer, C.:
\newblock Database encryption: an overview of contemporary challenges and
  design considerations.
\newblock SIGMOD Record \textbf{38}(3) (2009)  29--34

\bibitem{DBLP:reference/crypt/BouganimG11}
Bouganim, L., Guo, Y.:
\newblock Database encryption.
\newblock In: Encyclopedia of Cryptography and Security (2nd Ed.).
\newblock (2011)  307--312

\bibitem{DBLP:conf/stoc/Gentry09}
Gentry, C.:
\newblock Fully homomorphic encryption using ideal lattices.
\newblock In: Proceedings of the 41st Annual ACM Symposium on Theory of
  Computing, STOC 2009, Bethesda, MD, USA. (2009)  169--178

\bibitem{DBLP:conf/eurocrypt/BonehCOP04}
Boneh, D., Crescenzo, G.D., Ostrovsky, R., Persiano, G.:
\newblock Public key encryption with keyword search.
\newblock In: Proceedings of the Inteernational Conference on the Theory and
  Applications of Cryptographic Techniques (EUROCRYPT). (2004)  506--522

\bibitem{DBLP:conf/tcc/BonehW07}
Boneh, D., Waters, B.:
\newblock Conjunctive, subset, and range queries on encrypted data.
\newblock In: Proceedings of the 4th Theory of Cryptography Conference (TCC).
  (2007)  535--554

\bibitem{DBLP:conf/acns/BonehGHWW13}
Boneh, D., Gentry, C., Halevi, S., Wang, F., Wu, D.J.:
\newblock Private database queries using somewhat homomorphic encryption.
\newblock In: Applied Cryptography and Network Security - 11th International
  Conference, ACNS 2013, Banff, AB, Canada, Proceedings. (2013)  102--118

\bibitem{DBLP:journals/iacr/ZhengXA13}
Zheng, Q., Xu, S., Ateniese, G.:
\newblock Vabks: Verifiable attribute-based keyword search over outsourced
  encrypted data.
\newblock IACR Cryptology ePrint Archive \textbf{2013} (2013)

\bibitem{DBLP:conf/spw/BangerterCL04}
Bangerter, E., Camenisch, J., Lysyanskaya, A.:
\newblock A cryptographic framework for the controlled release of certified
  data.
\newblock In: Proceedings of the 12th International Workshop on Security
  Protocols. (2004)  20--42

\bibitem{DBLP:conf/idman/CamenischDLNPP13}
Camenisch, J., Dubovitskaya, M., Lehmann, A., Neven, G., Paquin, C., Preiss,
  F.S.:
\newblock Concepts and languages for privacy-preserving attribute-based
  authentication.
\newblock In: Policies and Research in Identity Management - Third IFIP WG 11.6
  Working Conference, IDMAN 2013. Proceedings. (2013)  34--52

\bibitem{DBLP:journals/eatcs/Gasarch04}
Gasarch, W.I.:
\newblock A survey on private information retrieval (column: Computational
  complexity).
\newblock Bulletin of the EATCS \textbf{82} (2004)  72--107

\bibitem{DBLP:conf/pairing/IovinoP08}
Iovino, V., Persiano, G.:
\newblock Hidden-vector encryption with groups of prime order.
\newblock In: Proceedings of the 2nd International Conference on Pairing-based
  Cryptography (Pairing). (2008)  75--88

\bibitem{DBLP:journals/joc/KatzSW13}
Katz, J., Sahai, A., Waters, B.:
\newblock Predicate encryption supporting disjunctions, polynomial equations,
  and inner products.
\newblock J. Cryptology \textbf{26}(2) (2013)  191--224

\bibitem{OkTa}
Okamoto, T., Takashima, K.:
\newblock Adaptively attribute-hiding (hierarchical) inner product encryption.
\newblock In: Proceedings of the Inteernational Conference on the Theory and
  Applications of Cryptographic Techniques (EUROCRYPT). (2012)  591--608

\end{thebibliography}


\end{document}